\documentstyle [12pt]{article}
\def\be{\begin{equation}}
\def\ee{\end{equation}}

\begin{document}

\title{Rigged Hilbert space of the free coherent states and $p$--adic numbers}
\author{S.V.Kozyrev}
\maketitle

\begin{abstract}
Rigged Hilbert space of the free coherent states
is investigated.
We prove that this rigged Hilbert space is isomorphous to the space
of generalized functions over $p$--adic disk.
We discuss the relation of the described isomorphism of rigged Hilbert spaces
and noncommutative geometry and show, that the considered example
realises the isomorphism of the noncommutative line and $p$--adic disk.
\end{abstract}

In the present paper, continuing the investigations of \cite{coherent1},
\cite{coherent2}, we investigate the free coherent states (or shortly FCS),
which are (unbounded) eigenvectors of the linear combination of
annihilators in the free Fock space.
In \cite{coherent1}, \cite{coherent2} it was shown that the space of
the free coherent states is highly degenerate for the fixed
eigenvalue $\lambda$ (and infinite dimensional), and this degeneracy
is naturally described by the space $D'(Z_p)$ of generalized functions
on $p$--adic disk ($p$ is a number of independent creators in the free Fock space).
In the present paper we reformulate the results of \cite{coherent1}, \cite{coherent2}
using the language of rigged Hilbert spaces and propose an interpretation
of the relation between the free coherent states and $p$--adics
using noncommutative geometry. We speculate that the isomorphism
between the space of FCS and the space of generalized function on $p$--adic
disk in the language of noncommutative geometry reduces to
the isomorphism between the noncommutative (or quantum) line and $p$--adic
disk.

$p$--Adic mathematical physics studies the problems of mathematical physics
with the help of $p$--adic analysis.
$p$--Adic mathematical physics
was studied in \cite{VVZ}--\cite{phase}.
For instance in the book \cite{VVZ} the analysis of $p$--adic
pseudodifferential operators was developed.
In \cite{Vstring} a $p$--adic approach in the string theory was proposed.
In \cite{Khren} a theory of $p$--adic valued distributions was investigated.
In \cite{ABK}, \cite{PaSu} it was shown that the Parisi matrix used
in the replica method is equivalent, in the simplest case,
to a $p$--adic pseudodifferential operator. In \cite{wavelets}
it was shown that the wavelet basis in $L^2(R)$ after the $p$--adic
change of variable (the continuous map of $p$--adic numbers onto real numbers
conserving the measure) maps onto the basis of eigenvectors of
the Vladimirov operator of $p$--adic fractional derivation.
In \cite{phase} a procedure to generate the ultrametric space
used in the replica approach was proposed.

The Free (or quantum Boltzmann) Fock space has been considered
in some works on quantum chromodynamics
\cite{MasterFld}--\cite{DougLi}
and noncommutative probability \cite{AccLu}--\cite{Speicher}.

The free Fock space ${\cal F}$ over a Hilbert space  ${\cal H}$
is the completion of the tensor algebra
$${\cal F}=\oplus_{n=0}^{\infty}{\cal H}^{\otimes n}.$$
Creation and annihilation operators act as follows:
$$
A^{\dag}(f) f_{1}\otimes \dots \otimes f_{n}=f\otimes
f_{1}\otimes \dots \otimes f_{n};\quad f,f_i \in {\cal H}
$$
$$
A(f) f_{1}\otimes \dots \otimes
f_{n}=\langle f,f_{1} \rangle f_{2}\otimes \dots  \otimes f_{n};
\quad f,f_i \in {\cal H}
$$
where  $\langle \cdot,\cdot \rangle $ is the scalar product
in the Hilbert space ${\cal H}$.
Scalar product in the free Fock space
(which we also denote $\langle \cdot,\cdot \rangle $)
is defined in the standard way.

In the case when  ${\cal H}$ is the $p$--dimensional complex Euclidean space
we have $p$ creation operators
$A^{\dag}_{i}$, $i=0,\dots ,p-1$;
$p$ annihilation operators
$A_{i}$, $i=0,\dots ,p-1$ with the relations
\begin{equation}\label{aac}
A_{i}A_{j}^{\dag}=\delta_{ij}.
\end{equation}
and the vacuum vector $\Omega$  in the free Fock space satisfies
\begin{equation}\label{vacuum}
A_{i}\Omega=0.
\end{equation}


The free coherent states (or shortly FCS) were introduced in
\cite{coherent1}, \cite{coherent2} as the formal eigenvectors of
the annihilation operator $A=\sum_{i=0}^{p-1}A_{i}$ in the free
Fock space ${\cal F}$ for some eigenvalue $\lambda$,
\begin{equation}\label{freecoherent}
A \Psi= \lambda \Psi.
\end{equation}
The formal solution of (\ref{freecoherent}) is
\begin{equation}\label{psi}
\Psi=\sum_{I} \lambda^{|I|} \Psi_I A^{\dag}_{I}\Omega.
\end{equation}
Here  the multiindex $I=i_0 \dots i_{k-1}$, $i_j \in\{0, \dots
,p-1\}$ and
\be\label{Adag}
A^{\dag}_{I}=A^{\dag}_{i_{k-1}} \dots
A^{\dag}_{i_0}
\ee
$\Psi_I$ are complex numbers which satisfy
\begin{equation}\label{cascade}
\Psi_I=\sum_{i=0}^{p-1}\Psi_{Ii}.
\end{equation}
The summation in the formula  (\ref{psi}) runs on all sequences
$I$ with finite length. The length of the sequence $I$ is denoted
by $|I|$ (for instance in the formula above $|I|=k$). The formal
series (\ref{psi}) defines the functional with a dense domain in
the free Fock space. For instance the domain of each free coherent
state for $\lambda\in (0,\sqrt{p})$ contains the dense space $X$
introduced below.

We define the free coherent state $X_I$ of the form
\begin{equation}\label{indicator}
X_I= \sum_{k=0}^{\infty} \lambda^k \left(\frac{1}{p}
\sum_{i=0}^{p-1}A_i^{\dag}\right)^k \lambda^{|I|} A^{\dag}_I
\Omega+ \sum_{l=1}^{\infty} \lambda^{-l}
\left(\sum_{i=0}^{p-1}A_i\right)^l \lambda^{|I|} A^{\dag}_I
\Omega
\end{equation}
The sum on $l$ in fact contains $|I|$ terms. For $\lambda\in
(0,\sqrt{p})$ the coherent state $X_I$ lies in the Hilbert space
(the correspondent functional is bounded).

We denote by $X$ the linear span of free coherent states of the form
(\ref{indicator}) and by $X'$ we denote the space of all the free
coherent states (given by (\ref{psi})).

Definition 1 was proposed and lemmas 2 and 7, corollary 4 and example 6 below
were proven in \cite{coherent1}, \cite{coherent2}.

\bigskip

\noindent{\bf Definition 1}.\qquad {\sl
We define the renormalized pairing of the spaces $X$ and $X'$ as follows:
\be\label{renormalizedproduct}
(\Psi,\Phi)=\lim_{\lambda\to\sqrt{p}-0}\left(1-\frac{\lambda^2}{p}\right)
\langle \Psi,\Phi\rangle \ee
Here $\Psi\in X'$, $\Phi\in X$.
}

\bigskip

Note that the coherent states $\Psi$, $\Phi$ defined by (\ref{psi}),
(\ref{indicator}) depend on $\lambda$ and the product $(\Psi,\Phi)$ does
not.

The correctness of the definition above is justified by the following lemma.

\bigskip

\noindent{\bf Lemma 2.}\qquad
{\sl
Vectors $X_I\in X$ lie in the domain of the functional $\Psi$ for $\lambda\in
(0,\sqrt{p})$ for an arbitrary free coherent state $\Psi$ defined by (\ref{psi}).
Moreover, the following limit exists and is equal to
\be\label{lemma2}
(\Psi,X_I)=\lim_{\lambda\to\sqrt{p}-0}\left(1-\frac{\lambda^2}{p}\right)
\langle \Psi,X_I\rangle=p^{|I|}\Psi_I
\ee
}

\bigskip

\noindent{\it Proof}\qquad
The pairing of the functional $\Psi\in X'$ given by (\ref{psi})
and the state (\ref{indicator})
is given by the following series
\begin{equation}\label{action}
\langle\Psi,X_I\rangle=\sum_{k=0}^{\infty}\lambda^{2k}
\langle\Psi^{k},X_I^{k}\rangle
\end{equation}
Here $\Psi^{k}$ and $X_I^{k}$ are the coefficients of $\lambda^k$
in the series for $\Psi$ and $X_I$.
$\Psi^{k}$ is defined by the formula
$$
\Psi^{k}=\sum_{|J|=k}  \Psi_J A^{\dag}_{J}\Omega;
$$
and $X_I^{k}$ for $k>|I|$ have the form
$$
X_I^k=
\left(\frac{1}{p} \sum_{i=0}^{p-1}A_i^{\dag}\right)^{k-|I|}
A^{\dag}_I \Omega.
$$
We obtain for $k>0$
$$
\langle\Psi^{|I|+k},X_I^{|I|+k}\rangle=
\langle X_I^{|I|+k},\Psi^{|I|+k}\rangle^{*}=
\langle X_I^{|I|+k-1},\frac{1}{p}\sum_{i=0}^{p-1}A_i\Psi^{|I|+k}\rangle^{*}=
$$
$$
=\langle X_I^{|I|+k-1},\frac{1}{p}\sum_{i=0}^{p-1}A_i
\sum_{|J|=|I|+k}  \Psi_J A^{\dag}_{J}\Omega\rangle^{*}
$$
This implies
$$
\frac{1}{p}\sum_{i=0}^{p-1}A_i
\sum_{|J|=|I|+k}  \Psi_J A^{\dag}_{J}\Omega=
\sum_{|J|=|I|+k-1}\frac{1}{p} \sum_{i=0}^{p-1} \Psi_{Ji} A^{\dag}_{I}\Omega=
\frac{1}{p}\Psi^{|I|+k-1}.
$$
Therefore
\be\label{chain}
\langle\Psi^{|I|+k},X_I^{|I|+k}\rangle=
\frac{1}{p}\langle\Psi^{|I|+k-1},X_I^{|I|+k-1}\rangle=
p^{-k}\langle\Psi^{|I|},X_I^{|I|}\rangle=p^{-k}\Psi_I
\ee
By (\ref{chain}) the series (\ref{action}) takes the form
$$
\langle\Psi,X_I\rangle=
\sum_{k=0}^{|I|}\lambda^{2k} \langle\Psi^{k},X_I^{k}\rangle
+\sum_{k=|I|+1}^{\infty}\lambda^{2k} p^{|I|-k}\Psi_I
$$
Since for $\frac{\lambda^2}{p}<1$ the series above is majorized by
the geometric series, the series (\ref{action}) converges and the
corresponding renormalized pairing takes the form
$$
(\Psi,X_I)=p^{|I|}\Psi_I
$$

This finishes the proof of the lemma.

\bigskip

Define the characteristic functions of $p$--adic disks
\begin{equation}\label{theta}
\theta_k(x-x_0)=\theta(p^{k}|x-x_0|_p);\quad
\theta(t)=0, t>1;\quad \theta(t)=1, t\le 1.
\end{equation}
Here $x$, $x_0\in Z_p$ lie in the ring of integer $p$--adic numbers
and the function $\theta_k(x-x_0)$ equals
to 1 on the disk $D(x_0,p^{-k})$ of radius $p^{-k}$ with the center in $x_0$
and equals to 0 outside this disk.

Identify multiindex $I=i_0\dots i_k$ with $p$-adic number
$I=\sum_{j=0}^{k}i_j p^j$.

The following lemma shows the relation between the renormalized pairing
of the free coherent states and the scalar product
of square integrable functions on $p$--adic disk.

\bigskip

\noindent{\bf Lemma 3}.\qquad {\sl
The space $X$ with the renormalized scalar product
is isomorphous, as a Euclidean space, to the space $D(Z_p)$ of test
functions on $p$--adic disk with the scalar product in $L^2$
with the isomorphism given by
\be\label{phi1}
\phi: X\to D(Z_p)
\ee
\be\label{phi2}
\phi: X_I\mapsto p^{|I|}\theta_{|I|}(x-I)
\ee
}

\bigskip

\noindent{\it Proof}\qquad
The space $X$ is a filtrated space with the filtration
$$
X=\bigcup_k X^{(k)},\qquad X^{(k+1)}\supset X^{(k)}
$$
where $X^{(k)}$ is generated by $X_I$ with $|I|=k$.
The spaces $X^{(k)}$ are finite dimensional.

Analogously, for $D(Z_p)$ there is the filtration by the finite dimensional
subspaces
$$
D(Z_p)=\bigcup_k D_k(Z_p),\qquad D_{k+1}(Z_p)\supset D_{k}(Z_p)
$$
where $D_{k}(Z_p)$ is generated by $\theta_k(x-I)$ with $|I|=k$.

To prove the lemma it is enough to prove that $\phi$ is the isomorphism
for the maps
$$
\phi: X^{(k)} \to D_{k}(Z_p)
$$

By the definition $D_{k}(Z_p)$ is a finite dimensional Euclidean space,
generated by the functions $\theta_{|I|}(x-I)$ with $|I|=k$.
The functions $\theta_{|I|}(x-I)$ obey the relation
\be\label{Drelat}
\sum_{i=0}^{p-1}\theta_{|Ii|}(x-Ii)=\theta_{|I|}(x-I)
\ee
and have the scalar products
\be\label{Dprod}
(\theta_{k}(x-I),\theta_{k}(x-J))=p^{-|I|}\delta_{IJ}
\ee

The space $X^{(k)}$ is generated by $X_I$ with $|I|=k$.
Vectors $X_I$ obey the following relation
\be\label{Xrelat}
X_I=p^{-1}\sum_{j=0}^{p-1} X_{Ij}
\ee
which we derive
from (\ref{indicator}) as follows:
$$
X_I= \sum_{k=0}^{\infty} \lambda^k \left(\frac{1}{p}
\sum_{i=0}^{p-1}A_i^{\dag}\right)^k \lambda^{|I|} A^{\dag}_I
\Omega+ \sum_{l=1}^{\infty} \lambda^{-l}
\left(\sum_{i=0}^{p-1}A_i\right)^l \lambda^{|I|} A^{\dag}_I
\Omega=
$$
$$
=p^{-1}\sum_{j=0}^{p-1} \sum_{k=0}^{\infty} \lambda^k \left(\frac{1}{p}
\sum_{i=0}^{p-1}A_i^{\dag}\right)^k \lambda^{|Ij|} A^{\dag}_{Ij}
\Omega + \lambda^{|I|} A^{\dag}_I\Omega+\sum_{l=1}^{\infty} \lambda^{-l}
\left(\sum_{i=0}^{p-1}A_i\right)^l \lambda^{|I|} A^{\dag}_I
\Omega=
$$
$$
=p^{-1}\sum_{j=0}^{p-1} \sum_{k=0}^{\infty} \lambda^k \left(\frac{1}{p}
\sum_{i=0}^{p-1}A_i^{\dag}\right)^k \lambda^{|Ij|} A^{\dag}_{Ij}
\Omega +
p^{-1}\sum_{j=0}^{p-1}\sum_{l=1}^{\infty} \lambda^{-l}
\left(\sum_{i=0}^{p-1}A_i\right)^l \lambda^{|Ij|} A^{\dag}_{Ij}
\Omega=
$$
$$
=p^{-1}\sum_{j=0}^{p-1} X_{Ij}
$$

Vectors $X_I$ have the following scalar products for $|I|=|J|$:
\be\label{Xprod}
(X_I,X_J)=p^{|I|}\delta_{IJ}
\ee

Comparing (\ref{Drelat}), (\ref{Dprod}) with (\ref{Xrelat}), (\ref{Xprod})
we obtain the statement of the lemma.

\bigskip

As a corollary, we obtain the following:

\bigskip

\noindent{\bf Corollary 4.}\qquad
{\sl
The renormalized scalar product
$( X_I,X_J )$ of the free coherent states  $X_I,X_J\in X$
equals to the integral over $p$--adic disk with respect to the Haar measure
$$
( X_I,X_J )=p^{|I|+|J|}
\int_{Z_p}\theta_{|I|}(x-I)\theta_{|J|}(x-J)\mu(dx)=
$$
\begin{equation}\label{scprodind}
=\left(
\frac{\theta_{|I|}(x-I)}{||\theta_{|I|}(x-I)||^{{2}}},
\frac{\theta_{|J|}(x-J)}{||\theta_{|J|}(x-J)||^{{2}}}
\right)_{L^2}
\end{equation}
}

\bigskip

\noindent{\bf Lemma 5.}\qquad
{\sl The isomorphism $\phi$ induces the injection $\phi'$ of the space $X'$
of free coherent states into the space $D'(Z_p)$ of generalized functions
over $p$-adic disk:
$$
\phi'(\Psi)=\Psi\circ\phi^{-1}
$$
}

\bigskip

\noindent{\it Proof}\qquad
We have to prove that for an arbitrary non--zero free coherent state $\Psi$
the functional $\Psi \circ \phi^{-1}$ is a
non--zero continuous linear functional over $D(Z_p)$.

By (\ref{lemma2}) and (\ref{phi2}) we have
\be\label{psioftheta}
(\Psi\circ \phi^{-1},\theta_{|I|}(x-I))=\Psi_I
\ee
Since for any non--zero coherent state $\Psi$ at least one coefficient $\Psi_I$
is non--zero, this proves that the functional $\Psi\circ \phi^{-1}$
is non--zero.

Topology in $D(Z_p)$ is defined as follows, see for instance \cite{VVZ}.
The space $D(Z_p)$ is
$$
D(Z_p)=\bigcup_{k}D_k(Z_p)
$$
where $D_k(Z_p)$ is a linear span of $\theta_{j}(x-I)$, $j\le k$ and the
sequence $\{\phi_j\}$ in $D(Z_p)$ is convergent when all $\phi_j\in D_k(Z_p)$
for all $j$ and some $k$ and the functions $\phi_j\to 0$ homogeneously.

Each of the spaces $D_k(Z_p)$ is the normed space (with the $C$--norm, equal
to the supremum over the $p$--adic disk of the modulus of the function).

Formula (\ref{lemma2}) implies that $\Psi\circ \phi^{-1}$
is a bounded functional on $D_k(Z_p)$ with the norm
$$
\|\Psi\circ \phi^{-1}\|_{D_k(Z_p)}={\hbox{ max }_{|I|\le k}}|\Psi_I|
$$
Therefore the functional $\Psi\circ \phi^{-1}$ is a continuous functional
on $D(Z_p)$, which finishes the proof of the lemma.

\bigskip

The important example of a generalized function is the $\delta$--function.
Let us introduce a coherent state that
correponds to the $\delta$--function.
Consider an infinite sequence $I=i_0 \dots i_{k}...$,
$i_j=0, \dots ,p-1$ and the
corresponding $p$--adic number $I=\sum_{k=0}^{\infty}i_k p^k$.
Let us denote $I_k=i_0 \dots i_{k-1}$.
We introduce the free coherent state $\delta_I$ of the form
$$
\delta_I= \sum_{k=0}^{\infty}
\lambda^{k} A^{\dag}_{I_k} \Omega.
$$

\bigskip

\noindent{\bf Example 6}.\qquad
{\sl The map $\phi'$ maps the free coherent state $\delta_I$ onto the
$\delta$--function:
$$
\phi'(\delta_I)= \delta(x-I)
$$
}

\bigskip

\noindent{\it Proof}\qquad Follows from (\ref{lemma2}) and (\ref{scprodind}).

\bigskip

\noindent{\bf Lemma 7}.\qquad
{\sl The injection of the space $X$
of free coherent states into the space $D'(Z_p)$ of generalized functions
over $p$--adic disk constructed in lemma 5 is surjective (and therefore
is an isomorphism of linear spaces).
}

\bigskip

\noindent{\it Proof}\qquad
To prove the lemma it is sufficient to construct the free coherent state
which, applied to the inverse image of the indicator of an arbitrary
$p$--adic disk in $Z_p$, will give an arbitrary complex number
(arbitrary up to the relation which follows from the linearity
of the functional and the fact that the indicator of the disk is equal
to the sum of the indicators of the subdisks).

This follows from the formula (\ref{psioftheta}):
$$
(\Psi, \phi^{-1}\theta_{|I|}(x-I))=\Psi_I
$$
where $\Psi_I$ is an arbitrary set of complex numbers satisfying
the relation (\ref{cascade}):
$$
\Psi_{I}=\sum_{i=0}^{p-1}\Psi_{Ii}
$$
which is exactly the property of linearity applied to the indicators of the
subdisks:
$$
(\Psi, \phi^{-1}\theta_{|I|}(x-I))=\sum_{i=0}^{p-1}
(\Psi, \phi^{-1}\theta_{|Ii|}(x-Ii))
$$
This finishes the proof of the lemma.

\bigskip

\noindent{\bf Corollary 8}.\qquad
{\sl The maps $\phi$, $\phi'$ (which are the isomorphisms of the linear spaces)
allow to map the topology of $D(Z_p)$ and $D'(Z_p)$ onto $X$ and $X'$
correspondingly. This procedure makes $\phi$ and $\phi'$ the
isomorphisms of vector topological spaces.
}

\bigskip

Lemmas 3, 5, 7 suggest the following definition.

\bigskip

\noindent{\bf Definition 9}.\qquad
{\sl
We denote $\tilde{\cal F}$ the completion of the space $X$ of the free
coherent states with respect to the norm defined by the renormalized
scalar product.
}

\bigskip

The space $\tilde{\cal F}$ is a Hilbert space with respect to the renormalized
scalar product.

\bigskip

\noindent{\bf Lemma 10}.\qquad
{\sl
The Hilbert space $\tilde{\cal F}$ lies in the space of the free coherent
states $X'$:
$$\tilde{\cal F}\subset X'$$
}

\bigskip

\noindent{\it Proof}\qquad
For $\Psi\in \tilde{\cal F}$, $\Psi=\lim_{n\to\infty}\Psi^{(n)}$,
$\Psi^{(n)}\in X$ consider the product $(\Psi,X_I)$ which we denote:
$$
(\Psi,X_I)=p^{|I|}\Psi_I
$$
Taking into account (\ref{Xrelat}) we have
$$
(\Psi,X_I)=
p^{-1}\sum_{i=0}^{p-1}(\Psi,X_{Ii})
$$
which implies that $\Psi_I$ satisfies (\ref{cascade}):
$$\Psi_I=\sum_{i=0}^{p-1}\Psi_{Ii}$$
Therefore
$$
\tilde\Psi=\sum_I \lambda^{|I|}\Psi_IA^{\dag}_I\Omega
$$
is the free coherent state in $X'$ with
$$
(\tilde\Psi,X_I)=(\Psi,X_I)=p^{|I|}\Psi_I
$$
which implies that $\tilde\Psi=\Psi$.
This finishes the proof of the lemma.

\bigskip

\noindent{\bf Lemma 11}.\qquad
{\sl The map
$$
j: X\to \tilde{\cal F}
$$
in (\ref{fcstriple}) is a continuous injection with the dense range.
}

\bigskip

\noindent{\it Proof}\qquad
Formula (\ref{lemma2}) implies that a non--zero coherent state
in $X$ has a non--zero norm in $\tilde{\cal F}$. Therefore $j$
is an injection.

Assume that $\{\Phi^{(n)}\}$ is a convergent (in the topology induced from
$D(Z_p)$) sequence in $X$. To prove the continuity of the injection $j$
we have to prove that the sequence $\{\Phi^{(n)}\}$ is fundamental in
$\tilde{\cal F}$.
By definition of the topology in $D(Z_p)$ there exists $k$ such that
$\{\Phi^{(n)}\}\subset X^{(k)}$. Therefore each $\Phi^{(n)}$
is a finite linear combination of the functions $X_I$, $|I|=k$,
and the convergence of the sequence $\{\Phi^{(n)}\}$ reduces to
the convergence of a finite number of coefficients in the decompostion over
$X_I$. By (\ref{lemma2}) this implies that $\{\Phi^{(n)}\}$
is a fundamental sequence in $\tilde{\cal F}$, which proves the continuity
of $j$.

This finishes the proof of the lemma.

\bigskip

Summing up the lemmas 10 and 11, we obtain the following:

\bigskip

\noindent {\bf Theorem 12}.\qquad {\sl
The space of the free coherent states
\be\label{fcstriple}
X  \stackrel{i}{\longrightarrow}  \tilde{\cal F}
\stackrel{j}{\longrightarrow}  X'
\ee
is a rigged Hilbert space.
}

\bigskip

Remind that a rigged Hilbert space is a triple of space
$$
A\stackrel{i}{\longrightarrow} {\cal H}\stackrel{j}{\longrightarrow} A^*
$$
where ${\cal H}$ is a Hilbert space, $A$ and $A^*$ are mutually conjugated
topological vector spaces, the maps $i:A\to {\cal H}$ and $j: {\cal H}\to
A^*$ are continuous injections, the image of $i$ is dense in ${\cal H}$,
and the maps $i$ and $j$ are conjugated in the following sense
$$
(ia,h)=(a,jh),\qquad a\in A, h\in {\cal H}
$$

We compare the rigged Hilbert spaces of the free coherent states (\ref{fcstriple})
and of generalized functions over $p$--adic disk:
$$
D(Z_p)\stackrel{i'}{\longrightarrow} L^2(Z_p)\stackrel{j'}{\longrightarrow} D'(Z_p)
$$

We arrive to the following theorem which is an extension of the theorem
proven in \cite{coherent2}.

\bigskip

\noindent {\bf Theorem 13}.\qquad {\sl The map $\phi$ defined by
$$
\phi:\quad X_I\mapsto p^{|I|}\theta_{|I|}(x-I);
$$
extends to an isomorphism $\phi$ of the rigged Hilbert spaces:

$$
\begin{array}{ccccc}
X & \stackrel{i}{\longrightarrow} & \tilde{\cal F} &
\stackrel{j}{\longrightarrow} & X' \\
\downarrow\lefteqn{\phi}&&\downarrow\lefteqn{\tilde\phi}&&
\downarrow\lefteqn{\phi'}\\
D(Z_p) & \stackrel{i'}{\longrightarrow} & L^2(Z_p) & \stackrel{j'}{\longrightarrow}
& D'(Z_p)
\end{array}
$$
between the rigged Hilbert space of the free coherent states (with the pairing given by the renormalized
scalar product)
and the rigged Hilbert space of generalized functions over $p$--adic disk
}

\bigskip

\noindent{\it Proof}\qquad The proof is by lemmas 3, 5, 7 and corollary 8.

\bigskip

\noindent {\bf Remark 14}.\qquad
Definition (\ref{freecoherent}) of the space of FCS:
$$
(A-\lambda)\Psi=0
$$
may be interpreted as the equation of the noncommutative (or quantum) plane $A=\lambda$. The free
coherent state $\Psi$ in this picture corresponds to a generalized function on a non--commutative
space (with non--commutative coordinates $A_i$, $A^{\dag}_i$) with support on the non--commutative
plane $A=\lambda$, $A=\sum_{i=0}^{p-1} A_i$.

The theorem 13 means that the space of generalized functions over the non--commutative plane is
isomorphic as a rigged Hilbert space to the space of generalized functions over a $p$--adic disk, or
roughly speaking the non--commutative plane is equivalent to a $p$--adic disk.

Let us note that
$$
\lambda=\sqrt{p}
$$
is the maximal possible value of $\lambda$ (the threshold). For $\lambda>\sqrt{p}$ any vector
(\ref{psi}) has an infinite norm and therefore does not lie in the Hilbert space.

\bigskip

\centerline{\bf Acknowledgements}

\smallskip

The author would like to thank I.V. Volovich for discussions and
valuable comments. This work has been partly supported by INTAS
(grant No. 9900545), CRDF (grant 10105),
and  The Russian Foundation for Basic Research
(project 02-01-01084).

\end{document}